\documentclass[aps,prl,twocolumn,groupedaddress,showpacs,floatfix]{revtex4}

\usepackage{graphicx}
\usepackage{float}
 \usepackage{bm}
 \usepackage{amsmath}
 \usepackage{natbib}
\begin{document}

\title{Out-of-equilibrium singlet-triplet Kondo effect in a single 
${\rm C}_{60}$ quantum dot}


\author{Nicolas Roch$^1$, Serge Florens$^1$, 
Vincent Bouchiat$^1$, Wolfgang Wernsdorfer$^1$ {\&} Franck Balestro$^1$
}

\affiliation{
  \textsuperscript{1}\,Institut N\'eel, CNRS {\&} Universit\'e J. 
  Fourier, BP 166,
38042 Grenoble Cedex 9, France}

\begin{abstract}
We have used an electromigration technique to fabricate a 
 $\rm{C_{{60}}}$ single-molecule transistor~(SMT). Besides describing 
 our electromigration procedure, we focus and present an
 experimental study of a single molecule quantum dot containing an 
 even number of electrons, revealing, for two different samples, a 
 clear out-of-equilibrium Kondo effect. Low temperature
 magneto-transport studies are provided, which 
demonstrates a Zeeman splitting of the finite bias anomaly.

\end{abstract}

\maketitle

\section{Introduction}
\label{intro}

\begin{figure*}
\includegraphics[width=16.5cm]{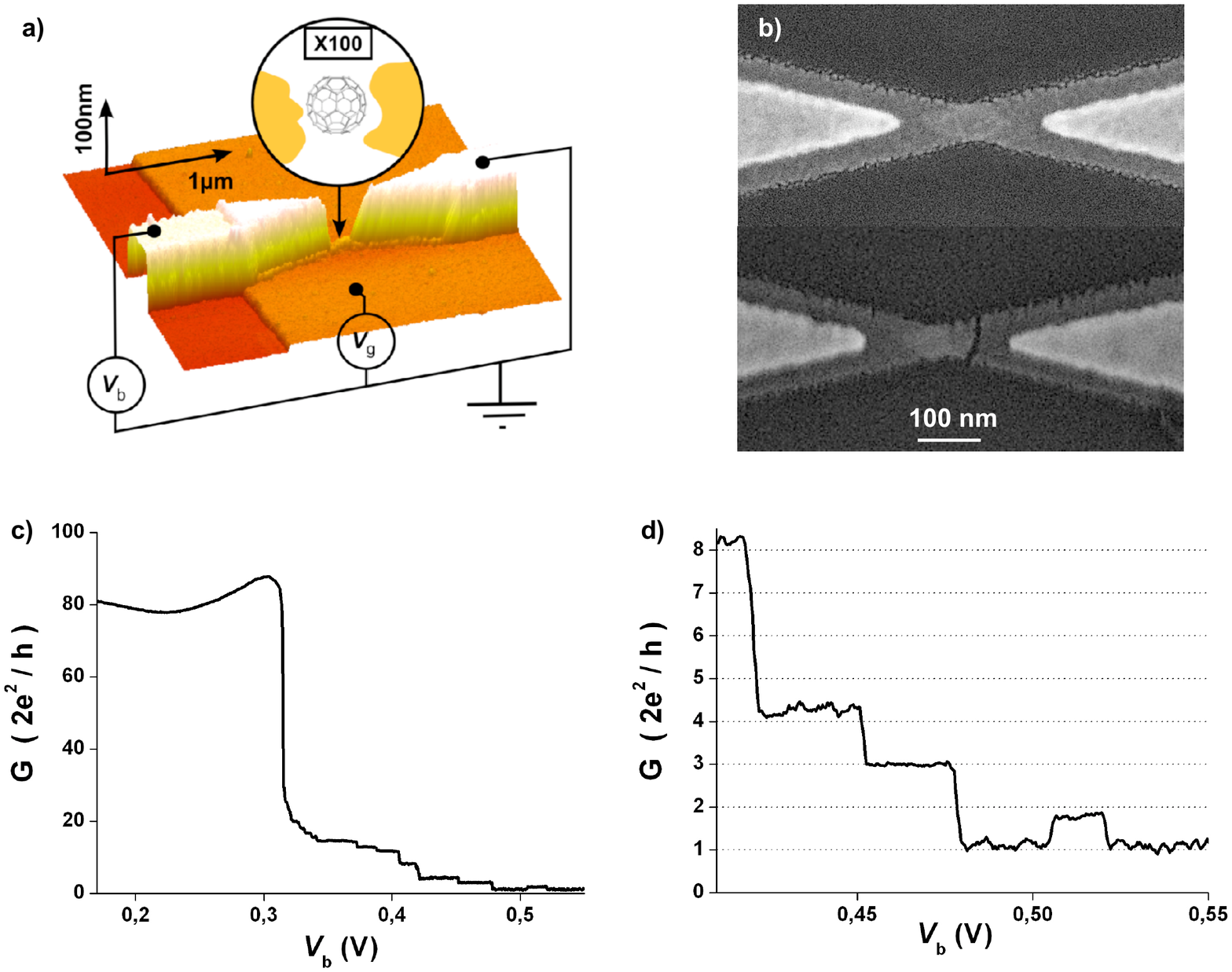}
\caption{ (color online) {\bf a)} Atomic-force-micrograph of the 
device : a gold nano-gap over an $Al/Al_{2}O_{3}$ back gate, with a 
single molecule trapped. {\bf b)} Field-emission SEM micrograph of 
the nanowire before and after the electromigration procedure. {\bf 
c)} Typical conductance trace obtained during the electromigration 
procedure, when the voltage is increased across the nanowire. {\bf 
d)} Quantized plateaus in the conductance before the opening of the 
nanogap.}
\label{electro}
\end{figure*}

Single-molecule transistor (SMT) is the ultimate limit in miniaturizating 
three terminal electronic devices, consisting in two reservoirs (Source and Drain) and
a gate (see Fig.~\ref{electro}a). While various experimental schemes have been proposed to 
address a single molecule quantum dot, one of the most attractive 
technique to realize a single-molecule transistor is the use of 
electromigration. This technique allows to realize nanometric gaps 
between electrodes by applying a controlled voltage ramp through a 
continuous metallic small wire. Due to the momentum transfer from the 
conduction electrons, and with some thermal enhancement due to Joule 
heating, metallic ions migrate and a nanogap between two electrodes 
is produced. If a single molecule is trapped during the 
electromigration process, due to 
nanometric confinement of the wave-function of the electrons in a 
SMT, Coulomb blockade phenomena are 
expected~\cite{McEuen2000,Ralph2002,Ward2008,Liang2002bis,Kubatkin2003,Champagne2005}. Depending
on the bias voltage $V_{\rm{b}}$ and the gate voltage 
$V_{\rm{g}}$, the transistor can be tuned to allow current flowing or not 
through the single-molecule, resulting in Coulomb diamond 
diagrams~(Fig.~\ref{fig3}a). The number of 
observed Coulomb diamonds in a given gate voltage regime strongly depends on the gate depending
coupling of $\rm{C_{{60}}}$ to surroundings and on the charging 
energy of the SMT.

In the presence of an odd number of electrons in a SMT, when a 
spin-1/2 magnetic state of the quantum dot is strongly coupled 
antiferromagnetically to the electrons in the reservoirs, the 
electronic states of the quantum dot 
hybridise with the electronic states of the reservoirs. As a result, 
even if the energy of the quantum dot state is far below the Fermi 
level of the reservoirs, hybridisation creates an effective 
density of states on the site of the dot, which is pinned at the Fermi level of the reservoirs, leading 
to a zero-bias anomaly where a Coulomb gap would have naively been 
expected. This is known as Kondo 
effect in quantum dot devices~\cite{Glazman1988,Ng1988}, and this signature has been
widely observed in semiconducting devices~\cite{Goldhaber1998,Cronenwett1998}, carbon
nanotube~\cite{Nygard2000},
or single-molecule~\cite{Liang2002bis,Yu2004,Roch2008} quantum dots.
Universality is a fundamental property of the Kondo effect and a single energy scale,
associated with the Kondo temperature $T_{\rm K}$, fully describes the
physical properties at low energy. When the typical energy of a
perturbation, such as temperature, bias voltage, or magnetic field,
is higher than $T_{\rm K}$, the coherence of the system is
suppressed and the Kondo effect disappears. However, a fundamental 
question is what happens when the quantum dot is occupied with an 
even number of electrons ? For a quantum dot with two electrons and two nearly degenerate orbital
levels, two different kinds of magnetic states occur: a singlet and a
triplet. We present then an experimental study of a single-molecule 
quantum dot with an even number of electrons and a singlet ground 
state. This SMT is strongly coupled to the 
electrodes, leading to an out-of-equilibrium singlet-triplet Kondo 
effect that we observed with two different $\rm{C_{60}}$ based SMT. First, we present the set-up we have developed to 
realize the electromigration procedure and the measurements, then we 
explain how the SMT is prepared, and finally, we present our measurements.

\section{Set-up description and electromigration procedure}
\label{sec:1}

\begin{figure*}
\includegraphics[width=15cm]{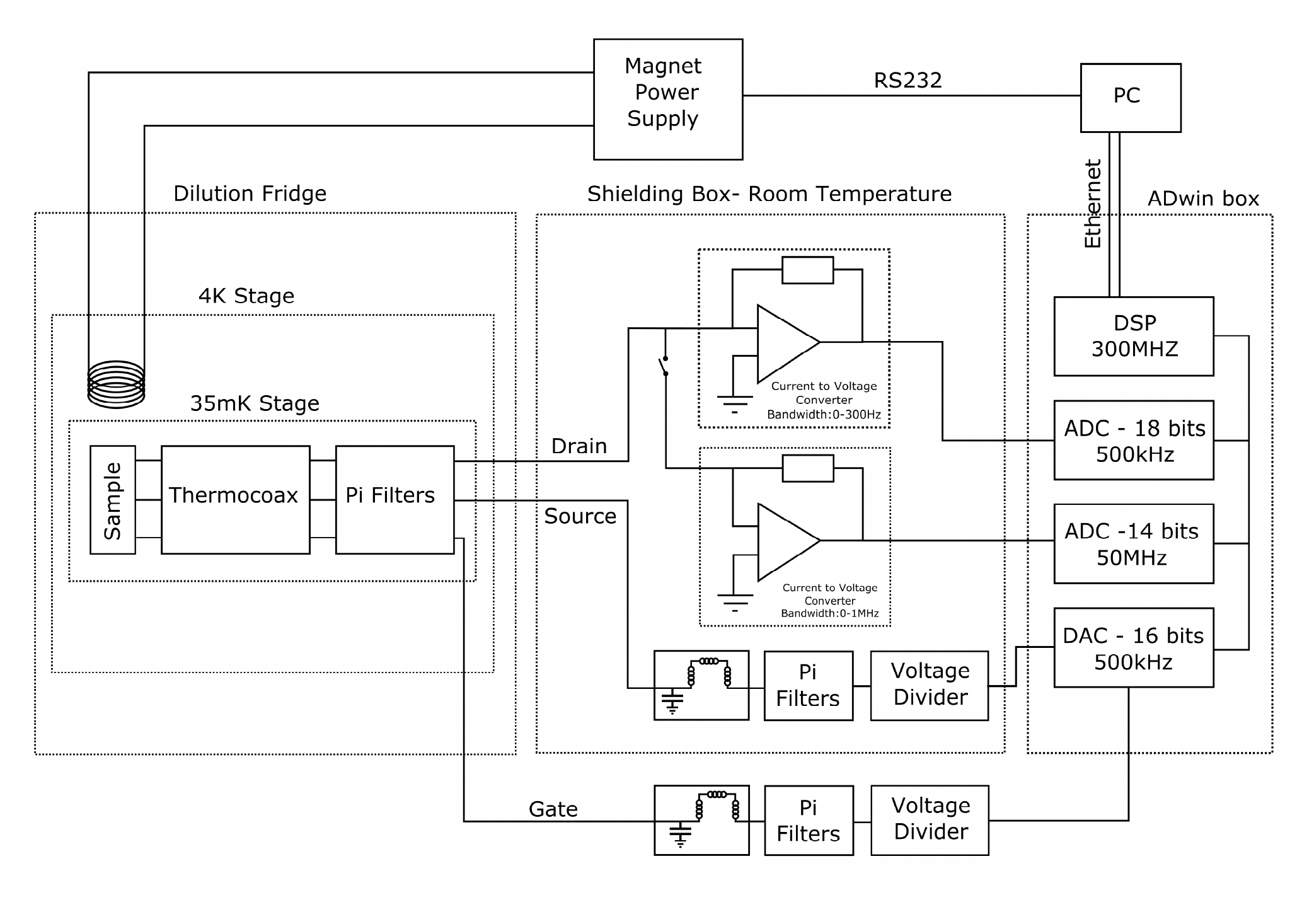}
\caption{ \label{figsetup} { Simplified scheme of the experimental
setup.} See text for details.}
\end{figure*}

To our knowledge, no electromigration 
procedure has been previously carried out in a dilution refrigerator with a 
high degree of filtering of the electrical leads. Indeed, the creation of nanogaps with this
technique requires minimizing the series
resistance~\cite{vanderZant2006}, which is generally incompatible
with dilution fridge wiring and filtering. To overcome this problem, we
developed a specific measurement setup (Fig.~\ref{figsetup}), which is 
divided into two parts, described here.

First, electromigration~\cite{Park1999} is performed at 4~K with the fast part of the
setup. As we wanted to perform such measurements in a dilution
fridge, we developed an efficient electromigration technique
since dilution wires and low-temperature filters are very resistive
and add an important series resistance to the sample (few hundreds
Ohms). Improvements of the original procedure~\cite{Park1999} have
already been reported recently~\cite{Strachan2005,Houck2005,Esen2005,Trouwborst2006,O'Neill2007,Wu2007}. We ramp the
voltage (typical rate is $10$ mV/s) across the gold nano-wire 
fabricated by shadow evaporation on an $\rm{Al/Al_{2}O_{3}}$ back 
gate (Fig.~\ref{electro}b), and measure 
its conductance using a very
fast feedback-loop~($\rm{1.5}$~$\rm{\mu s}$). As shown on 
Fig.~\ref{electro}c, the procedure can be divided into three parts. 
First, we measure a decrease of the conductance due to Joule heating of the 
nanowire. In a second part, the conductance increases, due to the 
melting of part of the nanowire. Finally, when the voltage becomes too 
high, namely when the gold ions migrate, the conductance suddenly 
drops to much lower values, quantized in $G_{0}=2e^{2}/h$, as shown on 
Fig.~\ref{electro}d. Using the fast feedback-loop, we set the voltage to
zero when the resistance exceeds a defined threshold, typically 
20~k$\Omega$. The fast
feed-back loop was achieved with real-time electronics (Adwin Pro
II) and a home-built high-bandwidth current to voltage converter, as
described in Fig.~\ref{figsetup}. With this technique, we obtained
small gaps (Fig.~\ref{electro}b) characterized by the tunnel
current measured after electromigration, without molecules, in 
previous experiments, resulting in a gap of the order of 1-2 nm.

The second step uses the low noise component of the setup to measure the
single-molecule transistor. In addition to low-temperature
filtering, we used ${\Pi}$ filters and ferrite bead filters developed at Harvard by
J. MacArthur and B. D'Urso~\cite{Marcus}. In order
to minimise
ground loops we integrated all the analogical electronics in a shielded box at
room temperature. Because of its great versatility, Adwin Pro II can
be programmed to perform DC or lock-in measurements, and apply gate
or bias voltages, thus minimizing the possibility of ground loops.
Depending on the measurements, we used an AC-excitation
between 3~$\rm{\mu V}$ and 100~$\rm{\mu V}$ for the lock-in technique.

\section{Preparation of the single-molecule transistor}

Preparation of the SMT was realized by blow drying
a dilute toluene solution of a $\rm{C_{60}}$ molecule onto a gold nanowire
realized on an $\rm{Al/Al_{2}O_{3}}$ back gate, see 
Fig.~\ref{electro}a for a
schematic view of the setup. Before blow
drying the solution, the electrodes were cleaned with acetone, ethanol,
isopropanol and oxygen plasma. The connected sample is
inserted in a copper shielded box,
enclosed in a high frequency low-temperature filter, anchored to the mixing chamber of
the dilution fridge having a base temperature equals to $35$~mK. The nanowire coated
with molecules is then broken by electromigration~\cite{Park1999},
via a voltage ramp at $4$~K.
As it is known that even if the 
electromigration procedure is 
well controlled, there is always a possibility to fabricate a few 
atoms gold 
aggregate~\cite{Houck2005} transistor, we studied several 
junctions prepared 
within the same procedure with a toluene solution only. 
In our opinion, it is relevant to state here that an "interesting" 
device to investigate must show at least one order of 
magnitude change in the current characteristics as a function of 
the gate 
voltage for a $1$~mV voltage bias, and a charging 
energy greater than $20$~meV. Within these restrictions, we tested 
$38$ bared junction 
with a toluene solution and $51$ with a dilute $\rm{C_{60}}$ toluene solution. 
While $3$ bared junction showed one order of magnitude changes in the current as a function of the 
gate voltage after electromigration, 
only $2$ had a charging energy higher than $20$~meV, and only $1$ of 
those $2$ exhibited a zero bias anomaly. These transport structures were also 
not very well defined.
For junctions prepared with a diluted $\rm{C_{60}}$ toluene solution, we 
measured $7$ 
junctions out of $51$ with one order of magnitude changes in the 
current as a function of 
gate voltage, and $6$ of those $7$ had a charging energy 
higher than $20$~meV and exhibited pronounced zero bias anomalies.
Here we report on an experimental study of transport measurements in a
single-molecule QD, as a function of bias voltage $V_{\rm{b}}$, gate voltage
$V_{\rm{g}}$, temperature $T$~($35$~mK $<$ $T$ $<$ $10$~K), and magnetic
field $B$ up to $8$~T on two different samples labelled $\bf{A}$ and 
$\bf{B}$.

\section{Out-of-equilibrium singlet-triplet Kondo effect}
\label{sec:2}

\begin{figure*}
\includegraphics[width=15.3cm]{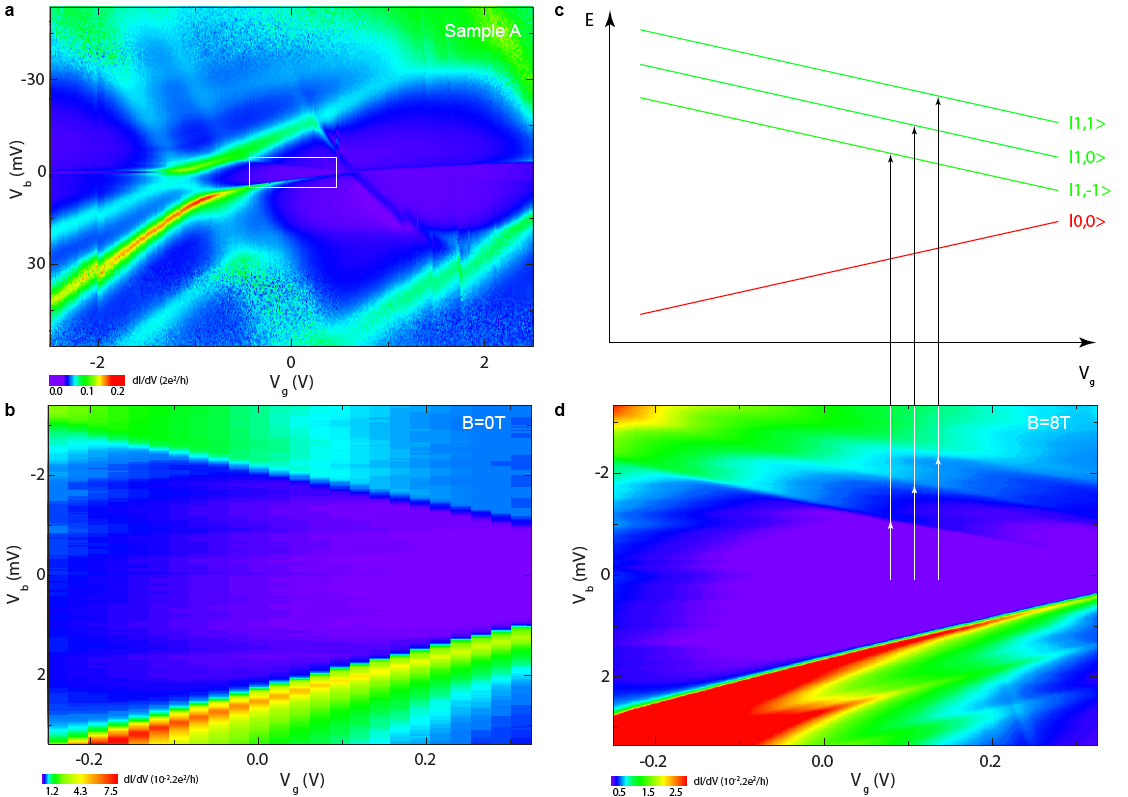}
\caption{ (color online) {\bf Transport and magneto-transport measurements for 
sample A :}
{\bf a)} Colour-scale map over two Coulomb diamonds of the differential conductance 
 $\partial I/\partial V$ as a function of bias voltage $V_{\rm{b}}$ and gate voltage 
 $V_{\rm{g}}$ at $35$~mK and zero magnetic field. 
 {\bf b)} Detailed characteristics of the differential conductance in 
 the region with an even number of electrons, corresponding to
 a low bias measurement inside the white rectangle in panel {\bf a)}.
 {\bf c)} Schematic of the singlet $\left|0,0\right>$ to 
 triplet $\left|1,-1\right>$, $\left|1,0\right>$ and 
 $\left|1,1\right>$ transitions at constant magnetic
 field.
 {\bf d)} $\partial I/\partial V$ measurements as a function of the 
 gate voltage $V_{\rm{g}}$,
 at fixed magnetic field $B=8$~T and temperature $T=35$~mK.}
\label{fig3}
\end{figure*}

In this section, we demonstrate, in our $\rm{C_{60}}$ molecular 
junction, an
effect recently reported by Paaske {\it{et al.}} in a carbon nanotube quantum
dot~\cite{Paaske2006}, namely the out-of-equilibrium singlet-triplet Kondo effect.
These authors were the first to clearly identify sharp finite
features as a Kondo effect and not as simple cotunneling via excited states.
The main idea behind Kondo physics is the existence of a degeneracy, which is
lifted by the conduction electrons. This is clearly the case for a quantum dot
with only one electron on the last orbital, leading to a doubly degenerate
spin $S=1/2$.

For a quantum dot with two electrons and two nearly degenerate orbital
levels, two different kinds of magnetic states occur: a singlet and a
triplet. Depending on $\delta E$, the energy difference between the
two orbital levels, and $J$, the strength of the ferromagnetic
coupling between the two electrons, the splitting between the triplet and
the singlet can in principle be tuned, and eventually brought to zero,
leading to the so-called singlet-triplet Kondo effect~\cite{Schmid2000}.
However the singlet is in most situations the ground state, leaving the triplet
in an excited state, thus suppressing the Kondo effect. Kondo signatures can
nevertheless be observed by tuning the degeneracy in a magnetic
field~\cite{Nygard2000,Sasaki2000,Liang2002}.

Another way to retrieve the degeneracy is to apply a bias voltage $V_{\rm{b}}$,
although it is of course more delicate to preserve the quantum coherence
necessary to Kondo correlations. Indeed, finite-bias features
clearly linked to magnetic excitations were observed in
2DEGs~\cite{Zumbuhl2004}, carbon 
nanotubes~\cite{Nygard2000,Liang2002,Babic2004,Quay2007} and even recently in an OPV5 molecule~\cite{Osorio2007}. 
However, only the study reported by Paaske {\it{et al.}}~\cite{Paaske2006}
was able to identify a clear out-of-equilibrium Kondo effect.
Their first observation was the occurence of sharp peaks in the differential
conductance for both positive and negative bias voltage, very different from
the cusps usually associated to cotunneling. Secondly the height of these peaks
decreased logarithmically with temperature, which is another typical signature of Kondo correlations.
Finally the shape of the peaks could be well accounted for in an out-of-equilibrium
Kondo calculation, while a simple cotunneling model failed to reproduce the
data.

\begin{figure*}
 \includegraphics[width=15.3cm]{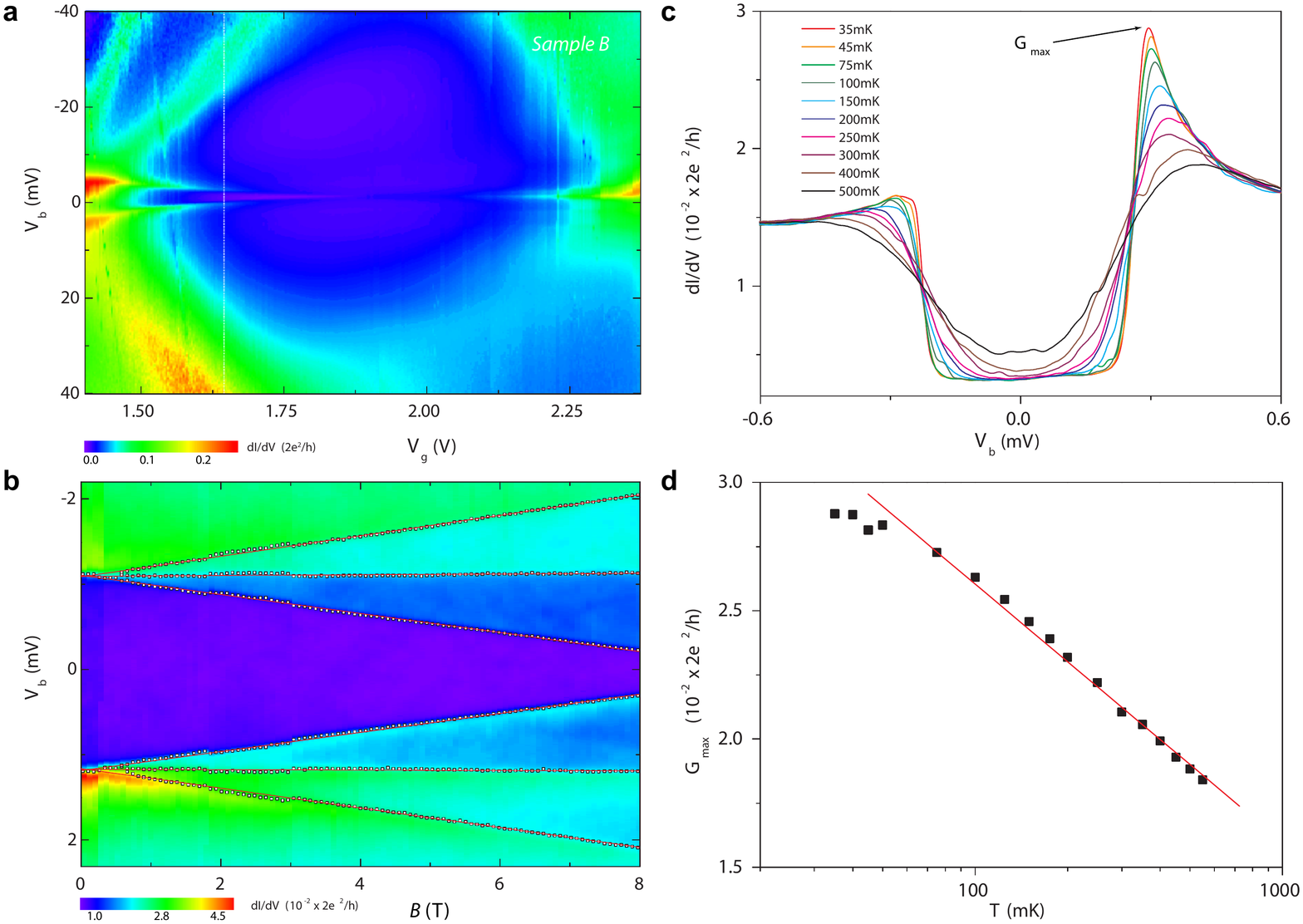}
 \caption{ (color online) {\bf Non-equilibrium singlet-triplet Kondo effect for 
 sample B.} {\bf a)}  Colour-scale map over one Coulomb diamond of the differential conductance 
 $\partial I/\partial V$ as a function of bias voltage $V_{\rm{b}}$ and gate voltage 
 $V_{\rm{g}}$ at $35$~mK and zero magnetic field. 
 {\bf b)} Differential conductance map as a function of bias voltage and
magnetic field at fixed $V_{\rm{g}}=1.64$~V. The linear fits in red demonstrate that
the non-equilibrium singlet-triplet Kondo peaks split at a finite
magnetic field $B_{\rm{c}}=50$~mT. 
{\bf c)} Differential conductance versus bias voltage for temperature from
35~mK~(pink) to 500~mK~(black) at fixed $V_{\rm{g}}=1.79$~V. 
{\bf d)} Evolution of the "positive $V_{\rm{b}}$" peak height in {\bf a)} with temperature on a
logarithmic scale, which can be linearly-fitted on nearly a decade.}
\label{fig4}
 \end{figure*}

These striking features are also present in our experiment, for the
case of an even number of electrons into the single $\rm{C_{60}}$ 
molecule for two different devices labelled sample {\bf A} and sample 
{\bf B}. These two-electron states can be described by their total spin 
$S$ and spin projection $m$ and are noted $\left|S,m\right>$. The
ground state of the system can thus be either a spin singlet $\left|0,0\right>$
with energy $E_{\rm{S}}$, or a spin triplet with energy $E_{\rm{T}}$ described by the three
states $\{\left|1,1\right>,\left|1,0\right>,\left|1,-1\right>\}$, 
degenerated at
zero magnetic field, but split by the Zeeman effect, with an energy shift
$\Delta E_{\rm{T}} = m g\mu_{\rm{B}}B$ for each state $\left|1,m\right>$, where ${g\approx{2}}$ 
for a $\rm{C_{60}}$ molecule.

Beginning with sample {\bf A}, we present in Fig.~\ref{fig3}a 
a colour-scale map over two Coulomb diamonds of the differential conductance 
 $\partial I/\partial V$ as a function of bias voltage $V_{\rm{b}}$ and gate voltage 
 $V_{\rm{g}}$ at $35$~mK and zero magnetic field. Focusing in the 
 right diamond with an even number of electrons, we present in 
 Fig.~\ref{fig3}b a precise 
 low-bias $\partial I/\partial V$ map corresponding to the white 
 rectangle in Fig.~\ref{fig3}a. In this region, we define, in 
 anticipation of our results, the singlet as the ground state of the 
 quantum dot. A gate-dependent finite-bias 
 conductance anomaly appears when $V_{\rm{b}}$ coincides with $E_{\rm 
 T}-E_{\rm S}$, which is due to an out-of-equilibrium Kondo effect 
 involving excitations into the spin-degenerate triplet. The possibility of gate-tuning the singlet-triplet 
splitting $E_{\rm T}-E_{\rm S}$ has already been demonstrated for 
lateral quantum dots~\cite{Kogan2003}, carbon nanotubes~\cite{Quay2007} and recently in
SMT~\cite{Roch2008}, and may originate from an asymmetric coupling of 
the molecular levels to the electrodes~\cite{Holm2007}. From now on we 
focus on the region where the singlet is the ground state, and 
present magneto-transport measurements to precisely identify these 
spin states.

Due to the high ${g\approx{2}}$ factor of a $\rm{C_{60}}$ molecule, 
as compared for instance to ${g\approx{0.44}}$ in GaAs-based devices, 
it is easier to lift 
the degeneracy of the triplet state via the Zeeman effect.
Fig.~\ref{fig3}d investigates the gate-induced out-of-equilibrium singlet-triplet 
Kondo effect at a constant magnetic field of 8~T. As the singlet is 
the ground state, the Zeeman split
triplet states are clearly measured as three parallel lines, which is in agreement with the
energy levels depicted in Fig.~\ref{fig3}c.

We now focus on sample {\bf B} and present first a colour-scale map 
over one Coulomb diamond, corresponding to a region with an even 
number of electrons, of the differential conductance 
 $\partial I/\partial V$ in Fig.~\ref{fig4}a. At a fixed gate value $V_{\rm{g}}$, 
 corresponding to the white line in a Fig.~\ref{fig4}a, we present a
$\partial I/\partial V$ measurement as a function of the 
magnetic field up to $8$~T in Fig.~\ref{fig4}b.
The splitting of the threefold triplet is, as for sample {\bf A}, 
clearly observed.
This plot, which was not numerically treated, shows the Zeeman splitting
between the three triplet states at both positive and negative bias.
The positions of those peaks are reported on
Fig.~\ref{fig4}b and a linear fit is applied to
each line, with a very good accuracy which enables us to determine, 
firstly, a
critical field $B_{\rm{c}}$ of 50~mT before the splitting occurs, and secondly, a Lande
factor $g=2\pm 0.1$. The existence of a critical field for the
splitting of the {\it zero-bias} anomaly is well-documented in the
case of the Kondo effect in equilibrium~\cite{Costi2000}, while more 
theoretical work is needed in order to interpret this data~\cite{Paaske2006}.

In Fig.~\ref{fig4}c, we present a
$\partial I/\partial V$ measurement as a function of the 
voltage bias $V_{\rm{b}}$, for a constant gate voltage $V_{\rm{g}}$, 
for different temperatures. While the conductance at low bias is suppressed when the spin state
of the system is a singlet, a clear finite-bias peak grows by
decreasing temperature as shown in
Fig.~\ref{fig4}c. In addition, the amplitude of
the positive bias peak decreases logarithmically about a decade
(Fig.~\ref{fig4}d), showing a clear signature of
the out-of-equilibrium singlet-triplet Kondo effect. 

To conclude, we have presented conductance, magneto-transport and 
temperature measurements on two different samples using 
an electromigration technique to realize a $\rm{C_{60}}$ based 
single-molecule quantum dot. These two SMTs clearly exhibit 
out-of-equilibrium singlet-triplet Kondo effect when the ground state 
of the quantum dot is defined by the singlet. In our opinion, it is 
extremely interesting, despite challenging, to study SMT because the
charging energy of a $\rm{C_{60}}$ molecule is possibly much larger
than that of a carbon nanotube quantum dot for example. Such systems 
allow to study new Kondo phenomena, such as quantum phase 
transition~\cite{Roch2008}, at relatively high temperatures.

We gratefully acknowledge E. Eyraud, D. Lepoittevin for their useful 
electronic and dilution technical contributions and motivating 
discussions. We thank E. Bonet, T. Crozes and T. Fournier for lithography 
development, C. Winkelmann, M. Deshmukh, T. Costi and L. Calvet for 
invaluable discussions. The sample of 
the investigations was fabricated in the NANOFAB facility of the Institut 
N\'eel.
This work is partially financed by ANR-PNANO Contract MolSpintronics. 


\end{document}